\documentclass[fleqn,usenatbib]{rasti}

\usepackage{newtxtext,newtxmath}

\usepackage[T1]{fontenc}

\DeclareRobustCommand{\VAN}[3]{#2}
\let\VANthebibliography\thebibliography
\def\thebibliography{\DeclareRobustCommand{\VAN}[3]{##3}\VANthebibliography}

\usepackage{graphicx}	
\usepackage{float}
\graphicspath{Figures/}
\usepackage[color=blue!20!white, obeyFinal]{todonotes}

\title[Learned interferometric imaging for radio interferometry]{Learned radio interferometric imaging for varying visibility coverage
}

\author[M. Mars et al.]{
Matthijs Mars,$^{1,2}$\thanks{E-mail: academic@matthijsmars.com}
Marta M.~Betcke,$^{3}$
Jason D.~McEwen$^{1,4}$\thanks{E-mail: jason.mcewen@ucl.ac.uk}
\\
$^{1}$Mullard Space Science Laboratory (MSSL), University College London (UCL), Dorking RH5 6NT, UK\\
$^{2}$Leiden Observatory, Leiden University, Leiden 2333 CC, The Netherlands\\
$^{3}$Department of Computer Science, University College London (UCL), London WC1E 6BT, UK \\
$^{4}$Alan Turing Institute, London NW1 2DB, UK
}

\date{Accepted XXX. Received YYY; in original form ZZZ}

\pubyear{2025}

\usepackage{xcolor}

\usepackage{soul}

\hypersetup{
  colorlinks=true,
}

\begin{document}
\label{firstpage}
\pagerange{\pageref{firstpage}--\pageref{lastpage}}
\maketitle

\begin{abstract}
  With the next generation of interferometric telescopes, such as the Square Kilometre Array (SKA), the need for highly computationally efficient reconstruction techniques is particularly acute. The challenge in designing learned, data-driven reconstruction techniques for radio interferometry is that they need to be agnostic to the varying visibility coverages of the telescope, since these are different for each observation. Because of this, learned post-processing or learned unrolled iterative reconstruction methods must typically be retrained for each specific observation, amounting to a large computational overhead. In this work we develop learned post-processing and unrolled iterative methods for varying visibility coverages, proposing training strategies to make these methods agnostic to variations in visibility coverage with minimal to no fine-tuning.  Learned post-processing techniques are heavily dependent on the prior information encoded in training data and generalise poorly to other visibility coverages. In contrast, unrolled iterative methods, which include the telescope measurement operator inside the network, achieve good reconstruction quality and computation time, generalising well to other coverages and require little to no fine-tuning. Furthermore, they generalise well to more realistic radio observations and are able to reconstruct images with with a larger dynamic range than the training set.
\end{abstract}

\section{Introduction}\label{sec:intro}
To study radio sources in the Universe at high-resolution there is a need for radio telescopes with a very large aperture that cannot be achieved with a single dish. 
Instead, radio interferometry forms a telescope using multiple dishes or antennas spaced out over long distances to synthesise an aperture large enough to image the Universe in enough detail to uncover its mysteries. 
By combining the measurements from two of these antennas, which is referred to as a baseline, a so-called visibility can be measured, corresponding to one Fourier component of the sky. 
The inversion of this measurement process is an ill-posed problem as the radio interferometric telescope samples the Fourier domain non-uniformly, with a finite number of samples, and are contaminated by measurement noise. 
By projecting the acquired data to the image domain one recovers what is called a dirty image, a reconstruction of the target convolved with the point-spread function (PSF) of the telescope. 
Typically, various algorithms are used to reconstruct the image from acquired visibilities to yield a more accurate clean image of the true radio sky \citep[e.g. CLEAN,][]{hogbomApertureSynthesisNonregular1974}.

However, processing the vast volume of data generated by current large radio interferometers, such as the LOw-Frequency ARray \citep[LOFAR,][]{vanhaarlemLOFARLOwFrequencyARray2013}, the Murchison Widefield Array \citep[MWA,][]{tingayMurchisonWidefieldArray2013}, and the Australian SKA Pathfinder \citep[ASKAP,][]{hotanAustralianSquareKilometre2021}, already presents immense computational challenges. 
The next generation of interferometric telescopes, such as the Square Kilometre Array \citep[SKA,][]{dewdneySquareKilometreArray2009}, promises to revolutionise radio astronomy by offering unparalleled sensitivity, resolution, and sky coverage across a large frequency range. 
With the even larger volume of data generated by the SKA, yielding terabytes of data per second, there is an ever rising need for image reconstruction techniques that are able to process this huge amount of data efficiently.

Current image reconstruction techniques such as CLEAN \citep{hogbomApertureSynthesisNonregular1974} and its multi-scale variants \citep{ bhatnagarScaleSensitiveDeconvolution2004, bhatnagarCorrectingDirectiondependentGains2008, stewartMultiplebeamCLEANImaging2011} can be computationally costly and also require supervision, provide suboptimal reconstruction quality, and cannot provide any uncertainty quantification.

In contrast, techniques based on compressive sensing have demonstrated high quality image reconstructions providing both sharp and smooth image features \citep[e.g.][]{carrilloSparsityAveragingReweighted2012,pratleyRobustSparseImage2018}. 
Highly distributed algorithms have been developed for these approaches to efficiently reconstruct images for large numbers of visibilities \citep[e.g.][]{pratleyDistributedParallelSparse2019} as well as to quantify the uncertainties \citep{caiUncertaintyQuantificationRadio2018a,caiUncertaintyQuantificationRadio2018}. 
Nonetheless, these methods need multiple applications of the measurement operator and its adjoint, which amounts to a large computational cost.

Recently, learned approaches have gained in popularity, using training data sets as prior information for reconstructions. One particular approach of these learned methods implements learned regularisation, replacing the conventional regularisers in optimisation methods with learned priors. 
The most popular and promising method is to use so-called Plug-and-Play \citep[PnP,][]{venkatakrishnanPlugandPlayPriorsModel2013} denoisers that can be trained on different data domains, foregoing the need for large domain-specific data sets. 
Since these methods retain the decoupling of the prior and measurement operator they are robust to changes in the measurement operator and measurement noise level. 
In particular, PnP priors have shown promising results in radio imaging \citep{terrisPlugandplayImagingModel2023, liaudatScalableBayesianUncertainty2024,terrisImageReconstructionAlgorithms2022,dabbechFirstAIDeep2022,wilberScalablePrecisionWidefield2023}.
Furthermore, recent work has demonstrated how PnP methods can be extended to provide efficient uncertainty quantification \citep{liaudatScalableBayesianUncertainty2024}. 
However, these methods require iterative reconstruction and so remain relatively computationally demanding.

Alternatively, one can use post-processing methods that use a denoising network to post-process an image created by using an approximate inverse of the measurement operator. 
In radio imaging this initial image is typically the dirty image and post-processing has been considered in several works \citep[e.g.][]{allamjrRadioInterferometricImage2016,terrisDeepPostProcessingSparse2019,ghellerConvolutionalDeepDenoising2021,connorDeepRadiointerferometricImaging2022,marsLearnedInterferometricImaging2023b}. 
Such post-processing networks depend more heavily on a correct model of the inverse of the measurement process and are more reliant on the prior information in the training set being representative of the actual observations. 
However, these methods are much faster than the learned regularisation methods as they only require a single evaluation of the measurement operator.

Learned unrolled iterative methods typically closely resemble iterative optimisation techniques.
These methods unroll a small, fixed number of iterations of the respective optimisation algorithm and replace the proximal operators with learned update steps, typically taking the form of a learned denoiser. 
Because these methods consist only of a few iterations, with the addition of learned updates, they are much faster than fully iterative approaches.
These methods have produced promising results in astronomical imaging \citep[e.g.][]{marsLearnedInterferometricImaging2023b,aghabiglouR2D2DeepNeural2024} and strike a balance between speed and quality of reconstruction. 
Inclusion of the measurement operator in the network architecture allows the network to retain some robustness to changes and uncertainty in the measurement operator \citep{boinkRobustnessPartiallyLearned2019, boinkPartiallyLearnedAlgorithmJoint2020, maierLearningKnownOperators2019}, however results are conditioned on the measurement operator used during training.

One essential challenge in radio interferometric imaging is the variability of the Fourier domain coverage of telescopes, which is referred to as the \emph{uv}-coverage. 
This is due to the coverage of the telescope depending on  factors such as the pointing of the telescope and the rotation of the earth. 
This variability in the \emph{uv}-coverage inevitably leads to a varying PSF for each observation and can therefore have a large impact on the quality of the reconstruction.
Since the measurement operator of the telescope changes for every observation, any approach reliant on the measurement operator in training would typically have to be retrained for each \emph{uv}-coverage.
While robustness to modest variations in the PSF for these methods has been investigated, e.g., by applying transformations to the PSF that model observational effects \citep{connorDeepRadiointerferometricImaging2022}, the use of post-processing and particularly learned unrolled iterative methods on different \emph{uv}-coverages is still outstanding.
Here we attempt to fill in this gap and study reconstruction methods that require minimal to no retraining for each observation by, e.g., training on a distribution of \emph{uv}-coverages or by using transfer learning to adapt the network to the specific \emph{uv}-coverage.

In this paper we discuss different strategies for training these learned reconstruction methods to determine how easily they can be adapted to the varying \emph{uv}-coverages. 
We compare methods that require no further training as well as methods that apply transfer learning with additional fine-tuning with only a small amount of retraining, compared to completely retraining the network for every different observation. 
We show that with only a small amount of overhead we can fine-tune these methods to the actual observations' \emph{uv}-coverages, leading to a significant improvement in reconstruction quality.

The remainder of this paper is structured as follows. 
In Section~\ref{sec:background} we discuss the interferometric imaging problem and how the measurement process can be modelled. 
Section~\ref{sec:learned-reconstruction-techniques} presents the different reconstruction approaches and the various training strategies that we develop for these learned methods. 
In Section~\ref{sec:results} we demonstrate the performance of the different training strategies on the reconstruction of our simulated data set as well as simulated measurements from an actual radio observation.
We discuss the implications of our results and the difficulties in applying these methods to real observations as well as how to potentially address these problems. 
Our findings are summarised in Section~\ref{sec:conclusion}.

\section{Radio interferometric imaging}\label{sec:background}
The monochromatic interferometric measurement process for a radio telescope is described as \citep[see, e.g.,][]{thompsonInterferometrySynthesisRadio2017}
\begin{equation}
  \begin{aligned}
    \mathcal{V} & (u, v, w) = \int_{-\infty}^{\infty} \int_{-\infty}^{\infty} \frac{1}{\sqrt{1-l^2-m^2}} A_N(l, m) I(l, m) \\
                & \times \exp \left\{-i 2 \pi\left[u l+v m+w\left(\sqrt{1-l^2-m^2}-1\right)\right]\right\} d l d m,
  \end{aligned}
\end{equation}
where $\mathcal{V}(u, v, w)$ is the complex visibility, $A_N(l, m)$ is the primary beam pattern, and $I(l, m)$ is the sky brightness distribution.
The coordinates $(u, v, w)$ depend on the coordinates of the baselines and the coordinates $(l,m)$ are directional cosines on the celestial sphere.
For telescopes with a small field-of-view the baselines are all approximately co-planar (i.e. $w=0$) and the measurement process reduces to the 2D Fourier transform
\begin{equation}\label{eq:visability}
  \mathcal{V}(u, v, 0)=\int_{-\infty}^{\infty} \int_{-\infty}^{\infty} \frac{A_N(l, m) I(l, m)}{\sqrt{1-l^2-m^2}} e^{-i 2 \pi(u l+v m)} d l d m.
\end{equation}

However for radio telescopes with a large field-of-view, the baselines are not co-planar and the measurement process is typically approximated by adding operations that account for the curvature by adding $w$-projection or $w$-stacking terms \citep[e.g.][]{cornwellNoncoplanarBaselinesEffect2008,tasseApplyingFullPolarization2013,offringaWSCleanImplementationFast2014, pratleyFastExactWstacking2019}.
In this work we limit ourselves to the co-planar case however the methods presented can be extended (at an increased computational cost) by adding these additional corrections.

\subsection{Interferometric measurement model}\label{sec:measurement-model}
The discretised version of Equation~\ref{eq:visability} can be modelled by a so-called non-uniform discrete Fourier transform (NUDFT) mapping from $N$ pixels in the observed image to $M$ non-uniformly distributed visibilities, with $\Phi: \mathbb{R}^N \rightarrow \mathbb{C}^M$ \citep{pratleyRobustSparseImage2018}.
The NUDFT can be approximated by the non-uniform fast Fourier transform \citep[NUFFT,][]{duijndamNonuniformFastFourier1997, jacksonSelectionConvolutionFunction1991, fesslerNonuniformFastFourier2003} which uses a gridding operation to interpolate the visibilities onto a regular grid and then uses the fast Fourier transform (FFT) to calculate the Fourier transform of the image.
The forward NUFFT operator can be expressed as a series of operators
\begin{equation}\label{eq:forward}
  \boldsymbol{\Phi } =   \boldsymbol{G} \boldsymbol{F} \boldsymbol{Z} \boldsymbol{D},
\end{equation}
where $\boldsymbol{D}:\mathbb{R}^N \rightarrow \mathbb{R}^N$ is the diagonal matrix that contains a correction in the image domain for the gridding operation in the Fourier domain, $\boldsymbol{Z}:\mathbb{R}^N \rightarrow \mathbb{C}^{\alpha^2 N}$ a zero-padding operation that pads and therefore enlarges the image by a factor of $\alpha$ in each direction to upsample the Fourier domain, $\boldsymbol{F}:\mathbb{C}^{\alpha^2 N} \rightarrow \mathbb{C}^{\alpha^2 N}$ is the unitary FFT, and $\boldsymbol{G}:\mathbb{C}^{\alpha^2 N} \rightarrow \mathbb{C}^M$ the degridding operation that uses an interpolation kernel to interpolate the visibilities from the regular grid to the non-uniform measurements.

The adjoint of the measurement operator, $\boldsymbol{\Phi}^*: \mathbb{C}^M \rightarrow \mathbb{R}^N$, can be expressed by using the adjoint of the individual operators in Equation~\ref{eq:forward} as
\begin{equation}\label{eq:adjoint}
  \boldsymbol{\Phi}^* = \boldsymbol{D} \boldsymbol{Z}^* \boldsymbol{F}^\dagger \boldsymbol{G}^*,
\end{equation}
with $\boldsymbol{G}^*: \mathbb{C}^M \rightarrow \mathbb{C}^{\alpha^2 N}$ the gridding operation that uses a gridding kernel to grid the measurements to the uniform grid, $\boldsymbol{F}^\dagger: \mathbb{C}^{\alpha^2 N} \rightarrow \mathbb{C}^{\alpha^2 N}$ the inverse unitary FFT, and $\boldsymbol{Z}^*: \mathbb{R}^{\alpha^2 N} \rightarrow \mathbb{R}^N$ the adjoint of the zero-padding operation, cropping the image back to its original size, and $\boldsymbol{D}^* = \boldsymbol{D}$ because of the self-adjointness of the real diagonal matrix correcting for the gridding operation.

To obtain an approximate inverse of the measurement process, we include measurement weights in the adjoint operation.
In radio interferometry the measurements are typically weighted using either natural, uniform or robust weighting schemes \citep{taylorSynthesisImagingRadio1999}.
Natural weights correspond to the inverse of the standard deviation of the measurement uncertainty $\boldsymbol{W}^{\text{Natural}}_{k,k} = \sigma_k^{-2}$, with $\sigma_k$ the uncertainty of the $k$-th visibility, $\mathcal{V}(u_k, v_k, 0)$, and is often also referred to as whitening.
Uniform weighting compensates for the increased sampling density at lower spatial frequencies by additionally weighing the measurements by the density of the sampling distribution, which causes the reconstructed image to be dominated by low spatial frequency power; $\boldsymbol{W}^{\text{Uniform}}_{k,k} = \sigma_k^{-2} N_s^{-1}(k)$, with $N_s(k)$ the number of measurements in a region around the measurement $\mathcal{V}(u_k, v_k, 0)$.
Lastly, robust weighting, also referred to as Briggs weighting, considers a trade-off between natural and uniform weighting, controlled by a robustness parameter \citep{briggsHighFidelityDeconvolution1995}.
The pseudo-inverse of the measurement operator applied to the measurements is typically referred to as the dirty image, and is calculated using
\begin{equation}\label{eq:pseudo-inverse}
  \boldsymbol{\Phi}^\dagger = \boldsymbol{D} \boldsymbol{Z}^* \boldsymbol{F}^\dagger \boldsymbol{G}^* \boldsymbol{W}.
\end{equation}
A more detailed explanation of the measurement process and the NUFFT operator can be found in, e.g.,  \citet{pratleyDistributedParallelSparse2019, marsLearnedInterferometricImaging2023b}.

\subsection{Inverse problem}
The radio interferometric measurement formation is given as
\begin{equation}
  \boldsymbol{y} = \boldsymbol{\Phi} \boldsymbol{x} + \boldsymbol{n},
\end{equation}
with $\boldsymbol{\Phi}: \mathbb{R}^N \rightarrow \mathbb{C}^M$ the measurement operator mapping from the true image $\boldsymbol{x} \in \mathbb{R}^N$ to the non-uniformly distributed measurements $\boldsymbol{y} \in \mathbb{C}^M$ with $\boldsymbol{n} \in \mathbb{C}^M$ the measurement noise.

The CLEAN algorithm solves the corresponding inverse problem using a greedy algorithm that aims to iteratively remove the brightest point sources from the observation in order to build up a clean sky model \citep{hogbomApertureSynthesisNonregular1974}.
Many variations of the algorithm exist that aim to improve or provide reconstruction of resolved, extended or polarised sources as well as multi-frequency synthesis \citep{clarkEfficientImplementationAlgorithm1980,schwabRelaxingIsoplanatismAssumption1984,steerEnhancementsDeconvolutionAlgorithm1984,saultApproachInterferometricMosaicing1996,cornwellMultiscaleCLEANDeconvolution2008,offringaWSCleanImplementationFast2014,pratleyImprovedMethodPolarimetric2016}.
These variants can be used to reconstruct images with a high dynamic range and are widely used in radio interferometry.
While the algorithm is very versatile, it has many parameters that need to be tuned for each observation and reconstructing using CLEAN is, by some, considered more an art than a science.

A popular alternative are variational approaches which formulate the inversion as a minimisation of the composite functional 
\begin{equation}
  \label{eq:variational-regularisation}
  \boldsymbol{x}^\star = \underset{\boldsymbol{x} \in \mathbb{R}^N}{\text{arg min}} \quad \mathcal{L}( \boldsymbol{\Phi} \boldsymbol{x}, \boldsymbol{y}) + \lambda \mathcal{S}(\boldsymbol{x}),
\end{equation}
which provides a trade off between the data fidelity term $\mathcal{L}(\boldsymbol{\Phi} \boldsymbol{x}, \boldsymbol{y})$ and the regularisation term $\mathcal{S}(\boldsymbol{x})$, which encodes prior information of the images and is weighted by the regularisation parameter $\lambda$.
In radio interferometry, a typical data fidelity term is the squared $\ell_2-\text{norm}$, i.e. $\mathcal{L}(\boldsymbol{\Phi} \boldsymbol{x}, \boldsymbol{y}) = \|\boldsymbol{\Phi} \boldsymbol{x} - \boldsymbol{y}\|_{\ell_2}^2$, and a widely used regularisation term is an $\ell_1$-norm on an overcomplete dictionary of wavelet bases, e.g. consisting of a Dirac basis and Debauchies wavelets bases \citep{daubechiesTenLecturesWavelets1992}, as used in the sparsity averaging reweighted analysis \citep[SARA,][]{carrilloSparsityAveragingReweighted2012} algorithm.

The optimisation problem in Equations~\ref{eq:variational-regularisation}
are typically solved using proximal optimisation techniques, such as the fast iterative shrinkage-thresholding algorithm \citep[FISTA,][]{beckFastIterativeShrinkageThresholding2009}, the alternating direction method of multipliers \citep[ADMM,][]{boydDistributedOptimizationStatistical2010}, or the primal-dual hybrid gradient method \citep[PDHG,][]{chambolleStochasticPrimalDualHybrid2018} and a variety of these methods have been applied to radio interferometry \citep[e.g.][]{pratleyRobustSparseImage2018,pratleyDistributedParallelSparse2019}.

Learned reconstruction approaches aim to replace some parts of these traditional optimisation techniques with neural networks in order to incorporate a data-driven prior and/or to speed up the reconstruction process.
In learned regularisation approaches neural networks capture the prior information in the variational problem. In particular, PnP methods replace the proximal problem formulation in proximal splitting algorithms with a denoiser \citep{venkatakrishnanPlugandPlayPriorsModel2013, ryuPlugandPlayMethodsProvably2019}.

\begin{figure}
  \centering
  \includegraphics[width=\columnwidth, trim= 0cm 0.5cm 0cm 1.1cm, clip]{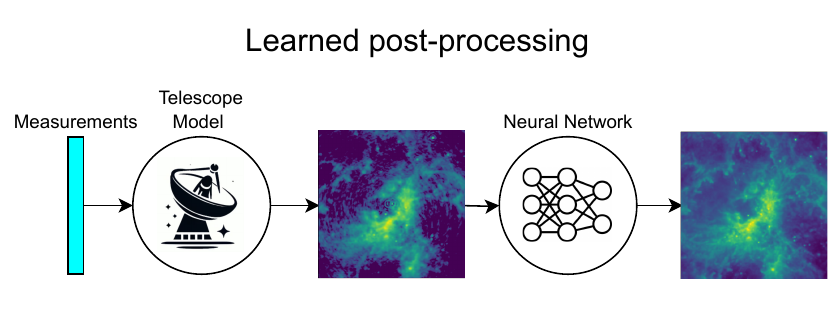}
  \caption{The learned post-processing approach creates a initial reconstruction from the measurements using a model of the telescope. The resulting dirty image is passed through a post-processing neural network to create the reconstruction. Our U-Net model takes this approach.}
  \label{fig:learned-post-processing-pipeline}
\end{figure}

Learned post-processing methods \citep[e.g.][]{chenLowDoseCTResidual2017,jinDeepConvolutionalNeural2017,yiSharpnessAwareLowDoseCT2018} aim to drastically speed up the reconstruction process by training a neural network to post-process an initial reconstruction.
Figure~\ref{fig:learned-post-processing-pipeline} shows the typical pipeline of a learned post-processing method in radio interferometry.
Because the network is decoupled from the measurement process, these methods can be trained using an image-to-image network making them much faster to train and evaluate.

\begin{figure}
  \centering
  \includegraphics[width=\columnwidth, trim= 0cm 0.4cm 0cm 1cm, clip]{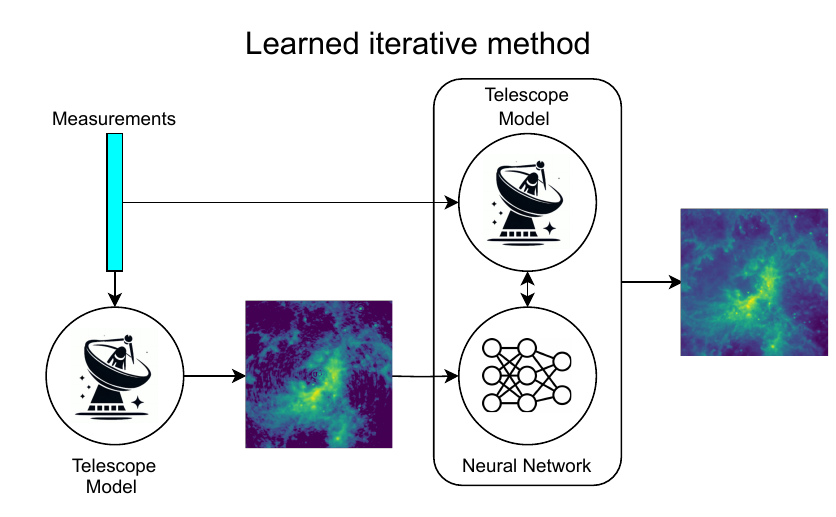}
  \caption{The learned unrolled iterative approach also creates an initial reconstruction using the telescope model and passes this to the reconstruction network. However, it also uses the measurements and the telescope model in conjunction with the neural network to update the image and provide the final reconstruction. Our GU-Net model takes this approach.}
  \label{fig:learned-unrolled-iterative-pipeline}
\end{figure}

Learned unrolled iterative methods combine the data-driven priors with model based updates and typically mimic proximal optimisation algorithms by unrolling a small number of iterations of the iterative process and replacing some of the proximal operators with neural networks \citep{gregorLearningFastApproximations2010}.
Figure~\ref{fig:learned-unrolled-iterative-pipeline} shows a diagram summarising these methods. 
While a large portion of these methods closely follow the traditional optimisation methods \citep[e.g.][]{adlerLearnedPrimaldualReconstruction2018}, some methods have been proposed that incorporate the measurement process more directly into the network.
For instance, \citet{hauptmannMultiScaleLearnedIterative2020} proposed a multi-scale learned iterative method that includes the measurement information at different resolution scales in the multi-scale U-Net architecture \citep{ronnebergerUNetConvolutionalNetworks2015}.
Similarly there are methods that see the inclusion of the gradient of a data fidelity term added into this U-Net structure at several resolution scales \citep{trentArchitectureInspiredSolvers2020,marsLearnedInterferometricImaging2023b}.
The benefit in these methods is the reduced computation and memory overhead as at the lower scales of the network the measurement operator is evaluated at a lower resolution and therefore more efficiently.

\section{Learned Reconstruction techniques}\label{sec:learned-reconstruction-techniques}
In order to provide fast image reconstruction of radio interferometric observations and to exploit the enhanced expressiveness of learned data-driven priors, we propose to use either learned post-processing or learned unrolled iterative methods.
These approaches minimise the number of applications of the measurement operator and its adjoint, which is the dominating factor in the computational cost of the reconstruction.
In particular, we consider the approaches introduced in our prior work \citep{marsLearnedInterferometricImaging2023b}, where these learned methods were used to provide efficient reconstruction for an optical interferometric telescope called SPIDER.
While in that case the \emph{uv}-coverage was fixed we now discuss how these methods can be adapted to work with varying \emph{uv}-coverages, via different training strategies tailored towards \emph{uv}-coverage generalisation.
We expect the unrolled iterative method to be more robust to changes in the \emph{uv}-coverage as it includes the measurement operator in the network architecture, while the post-processing method is more reliant on the prior information in the training data being representative of the actual observations.

\subsection{Learned post-processing method}
\begin{figure*}
  \centering
  \includegraphics[width=\textwidth, trim= 0 12cm 5cm 0, clip]{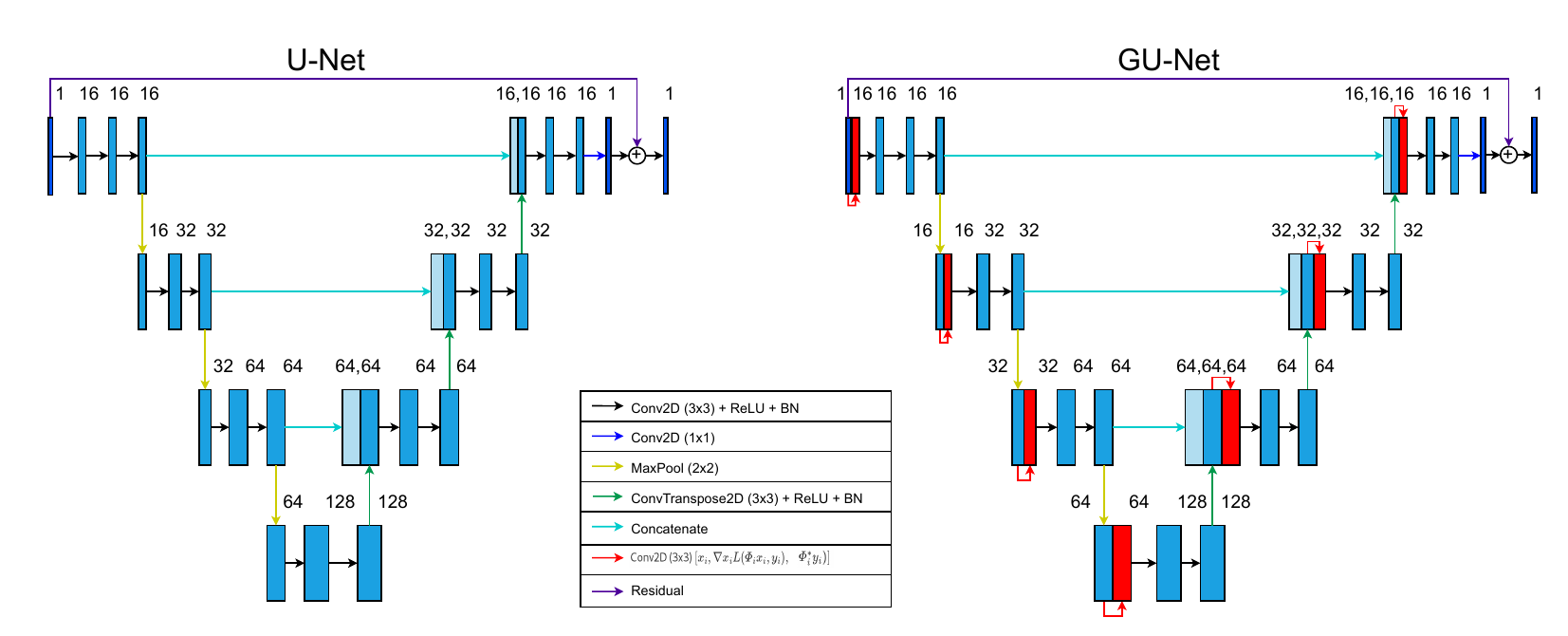}
  \caption{ Left: The U-Net architecture for the learned post-processing approach, with the input of the model being the dirty image. Right: The GU-Net architecture for the learned unrolled iterative approach that incorporates both the measurements and the measurement operator in the network through the (sub-scale) gradients. The model-based update takes the image in the first channel and calculates its gradient (Equation~\ref{eq:gradient}) and the sub-sampled dirty image (Equation~\ref{eq:dirty}), combines them using a convolutional layer, and concatenates this with the original channels. }
  \label{fig:networks}
\end{figure*}

The first reconstruction approach we consider is a learned post-processing approach that takes the dirty reconstruction as input and uses a U-Net architecture to remove artifacts from this initial reconstruction.
A schematic representation of the network architecture can be found in Figure~\ref{fig:networks}. The reconstruction can be formulated as:
\begin{equation}\label{eq:post-processing}
  \boldsymbol{x}^\star = \boldsymbol{\Phi}^\dagger_\theta \boldsymbol{y} = \boldsymbol{\Lambda}_\theta \boldsymbol{\Phi}^\dagger \boldsymbol{y},
\end{equation}
where $\boldsymbol{\Lambda}_\theta: \mathbb{R}^N \rightarrow \mathbb{R}^N$ denotes the reconstruction network and $\boldsymbol{\Phi}^\dagger: \mathbb{C}^M \rightarrow \mathbb{R}^N$ the pseudo-inverse of the measurement operator using natural weighting.
This post-processing is computationally efficient since the measurement operator is evaluated only once to create the initial reconstruction.
However, the final reconstruction quality is limited by the quality of this initial reconstruction.
Besides this, these reconstruction networks will learn how to remove errors that correspond specifically to the \emph{uv}-coverage used in the training data.
The implementation of the U-Net used in this article is similar to that of the U-Net used in \citet{marsLearnedInterferometricImaging2023b}, with the addition of a residual connection between the input and output of the network.
The initial reconstructions are created using the appropriate measurement operator for the \emph{uv}-coverage of the observation.

\subsection{Learned unrolled iterative method}
The second approach we consider is a learned unrolled iterative method that uses the measurement operator and measurements of the observation in the neural network to improve the reconstruction quality.
We extend the U-Net architecture by introducing measurement information at each scale of the network.
To achieve this we use so-called sub-scale operators $\boldsymbol{\Phi}_i: \mathbb{R}^N_i \rightarrow \mathbb{C}^M_i$ that map from a reduced image space $\boldsymbol{x}_i \in \mathbb{R}^{N_i}$ to low-pass filtered visibilities $\boldsymbol{y}_i \in \mathbb{C}^{M_i}$.
These are applied to the down-sampled images at each scale of the network. The Fourier domain of these sub-scale operators is restricted by applying a low-pass filter to the Fourier domain of the full measurement operator, such that the sub-scale operators only contain the measurement information that is relevant for the current scale of the images.
The choice of the low-pass filter is discussed in more detail in \citet{marsLearnedInterferometricImaging2023b}.
These sub-scale operators are used to include the gradient of the data fidelity term $\mathcal{L}( \boldsymbol{\Phi}_i\boldsymbol{x}_i, \boldsymbol{y}_i) = \frac{1}{2} \| \boldsymbol{\Phi}_i\boldsymbol{x}_i - \boldsymbol{y}_i\|_{\ell_2}^2$ in the multi-scale U-Net:
\begin{equation}\label{eq:gradient}
  \nabla_{\boldsymbol{x}_i} \mathcal{L}( \boldsymbol{\Phi}_i\boldsymbol{x}_i, \boldsymbol{y}_i) \propto  \boldsymbol{\Phi}_i^*(\boldsymbol{\Phi}_i\boldsymbol{x}_i- \boldsymbol{y}_i),
\end{equation}
where $\boldsymbol{x}_i \in \mathbb{R}^{N_i}$ is the current iterate at scale $i$ of the network and $\boldsymbol{y}_i \in \mathbb{C}^{M_i}$ are the low-pass filtered visibilities at scale $i$ of the network.

In addition to the gradient information we also include the sub-sampled dirty image, $\boldsymbol{x}_{i,\text{dirty}} \in \mathbb{R}^{N_i}$ at each of the gradient update steps
\begin{equation}\label{eq:dirty}
  \boldsymbol{x}_{i,\text{dirty}} = \boldsymbol{\Phi}^\dagger_i\boldsymbol{y}_i.
\end{equation}

We call the resulting network a Gradient U-Net \citep[GU-Net,][]{marsLearnedInterferometricImaging2023b}.
While this network includes several gradient updates (in this case seven), only two of these are at the full image scale and the rest are at the sub-scales of the network, which can be computed at a much lower computational cost because of the reduction in image size and restriction of the Fourier space.

In contrast to our previous work \citep{marsLearnedInterferometricImaging2023b}, we do not include a filtered gradient, corresponding to the gradient of a weighted data fidelity term, in the network architecture as its contribution is marginal compared to the increase in computational time and thus training time that it requires.
We also include a residual connection between the input and output of the network.
The GU-Net architecture is displayed in Figure~\ref{fig:networks}.


\subsection{Training strategies}\label{sec:training-strategies}
In order to train a reconstruction network that is robust against the variation in the \emph{uv}-coverage of the radio telescope, we propose several different training strategies.
In these strategies we differentiate between the \emph{true coverage}, which is the \emph{uv}-coverage of the observation we are trying to reconstruct, and other coverages drawn from a similar distribution as the true coverage, yet different to the actual observation.

The \emph{uv}-coverages are drawn from a Gaussian distribution, yet the training strategies could work with coverages drawn from a distribution corresponding to a specific telescope in the future.
We also discuss the results of fine-tuning a model to a realistic \emph{uv}-coverage from the MeerKAT telescope using transfer learning in Section~\ref{sec:transfer-meerkat}.

\subsubsection{Training on the true coverage}
The first training strategy is to train networks on the true coverage of the observation.
The benefit of this strategy is that it is conditioned on the exact \emph{uv}-coverage of the observation, yet this can only be known after the observation.
Therefore, the networks would have to be fully trained after each observation.
This strategy can be seen as the "oracle" strategy as it is the best case scenario if you have perfect knowledge of the observation.
We will use this strategy as a best case benchmark to evaluate the performance of the other strategies.

\subsubsection{Training on a single coverage}
Instead of training on the true coverage, we can also use training data that is generated with a single alternate coverage that is drawn from a distribution of coverages representative of the radio telescope.
At time of actual reconstruction, the measurement operators that were used during training can be swapped out for operators that use the true coverage of the observation.
This approach is expected to perform better using reconstruction approaches that rely less on the prior information related to the measurement model in the training data and that leverage the measurement operator in the reconstruction process, such as the GU-Net approach.

\subsubsection{Training on a distribution of coverages}\label{sec:distribution-strategy}
Alternatively, we can learn a reconstruction approach that is conditioned on the distribution of coverages representative of the radio telescope.
In order to do so, we create training data using a different random \emph{uv}-coverage for every batch to train our networks.
The performance of this method depends on how well the true coverage is characterised by the distribution of coverages used for training.
If there is a large variance in the distribution of coverages, the model will have a harder time learning the underlying connection between the measurements and the image.

Training a post-processing method using this strategy is trivial as the network can be trained as an image-to-image network without explicit calls to the measurement operator in the training process.
However, for the learned unrolled iterative method, the measurement operator is explicitly included in the training process, posing a challenge as it needs to be varied throughout training.
In order to achieve this there are three possible approaches.
The first option is to swap the measurement operator in the model during training, which is not supported by most deep learning frameworks and therefore requires recompiling the execution graph whenever the operator needs to be swapped, resulting in significantly increased training times.
For the second option, the \emph{uv}-coverage can be added as an input to the network in order to create the measurement operator as part of the training process, though this results in a large computational overhead in order to compute the coefficients for the (de)gridding of the non-uniformly distributed visibilities on the fly.
For the third option, we can compile the network with all possible measurement operators and use a mask to select the correct operator for the current batch of measurements.
In general, this results in memory overheads because of the large number of operators that need to be compiled into the network.
For the case of the random Gaussian coverages, we use this third approach, generating a larger number of visibilities and randomly select a sub-selection of these for each batch of training.
In this way, the memory overhead is reduced as the random visibilities will share some visibility measurements between them.

\subsubsection{Transfer learning models}
While the single coverage and distribution of coverages strategies do not depend on the true coverage and therefore do not need any retraining at observation, we can use transfer learning to fine-tune these strategies.
This is achieved by using the weights of the models that were trained using these strategies as a starting point and fine-tuning these weights by training the model using a data set created using the true coverage.
Since the number of training epochs needed to fine-tune the models is much smaller than the number needed to fully retrain the models, this results in a much more manageable training process.
We can also forego the use of data augmentation as the risk of overfitting is smaller.
The specifics of the data sets used for training and transfer learning as well as data augmentation are discussed in Section~\ref{sec:training-data}.
This results in reconstruction networks adapted to the specific \emph{uv}-coverage of the observation at a relatively modest cost of retraining for each observation.

\subsection{Training and training data}\label{sec:training-data}
To train the learned reconstruction networks we create a data set of measurements using images of simulated galaxies created from the IllustrisTNG simulations \citep{nelsonFirstResultsTNG502019, nelsonIllustrisTNGSimulationsPublic2019}, which were obtained by binning and in-painting the image from the simulated particles as described in more detail in \citet{marsLearnedInterferometricImaging2023b}.
The images are augmented during training by randomly flipping and rotating them. We use an NUFFT operator to simulate radio interferometric measurements, as discussed in Section~\ref{sec:measurement-model}, from the $256\times256$ images, using an upsampling factor of 2 and Kaiser-Bessel kernels with a size of $6\times6$ pixels.
The \emph{uv}-coverage used is drawn from a 2D Gaussian distribution and results in a total of $32768$ measurements.
These measurements are contaminated with complex Gaussian noise with standard deviation
\begin{equation}
  \sigma = \frac{\| \boldsymbol{\Phi} \boldsymbol{x}\|_{\ell_2}}{\sqrt{M}} \cdot 10^{\frac{-\text{ISNR}}{20}},
\end{equation}
with the input signal-to-noise ratio (ISNR) of 30dB, and this noise is re-drawn for every epoch of training.

From the IllustrisTNG images, 2000 images are used for training and 1000 for testing.
These image sets are sampled with the \emph{uv}-coverages drawn from the same Gaussian distribution to create several different training sets.
First, we create a train and test data set based on the true \emph{uv}-coverage that will be used to train the "oracle" network and to evaluate the performance of the different models.
Then, a second different \emph{uv}-coverage is used to create a training data set for the single coverage model.
Finally, a set of random \emph{uv}-coverages is used to create a data set to train the distribution of coverages model.
For this data set, every batch of training data is generated using a randomly generated coverage.
The data set used for the transfer learning of the models uses the true coverage to create measurements, but does not use any type of data augmentation as considered for the previous data sets, such that the input-output pairs for the model only have to be generated once for the full training of these models, keeping the time needed for training low.

The networks are trained for 1000 epochs using the Adam optimizer \citep{kingmaAdamMethodStochastic2017} with a learning rate of $10^{-3}$ and a batch size of 20, using a mean squared error loss between the true image and the reconstructed image.
Transfer learning using the non-augmented data set based on the true coverage is performed for 100 epochs.
The data sets and the weights of the networks are 32-bit floating point numbers and the networks are trained on a single NVIDIA A100 GPU.
\section{Results of numerical experiments}\label{sec:results}
We evaluate the effectiveness of the proposed training strategies in terms of reconstruction quality, (re)training times, robustness to variation within the measurement distribution and ability to generalise to more realistic radio observations.
First, we show that using these strategies reduces the time necessary to prepare a model for each observation.
The quality of reconstructions is assessed on images from the IllustrisTNG data set and an example of a realistic radio observation in terms of the signal-to-noise ratio, $\text{SNR}(\boldsymbol{x}_{\text{true}}, \boldsymbol{x}_{\text{pred}}) = 20 \cdot \log_{10} \left( {\frac{\lVert \boldsymbol{x}_{\text{true}} \rVert }{\lVert \boldsymbol{x}_{\text{pred}} - \boldsymbol{x}_{\text{true}} \rVert }}\right)$ with respect to the true image.
For the more realistic radio observations, we also evaluate the reconstructions on a logarithmic scale taking into account their extended dynamic range.
For these evaluations we track the logarithmic SNR; $\text{logSNR}(\boldsymbol{x}_{\text{true}}, \boldsymbol{x}_{\text{pred}}) = \text{SNR}(\log_{10}(\boldsymbol{x}_{\text{true}}), \log_{10}(\boldsymbol{x}_{\text{pred}}))$.
Finally, we demonstrate how these models can be adapted for use with actual radio \emph{uv}-coverages by transferring the learned models to a real telescope and evaluating their performance on a realistic radio observation.

\subsection{Training and evaluation times}
\begin{figure}
  \centering
  \includegraphics[width=\columnwidth]{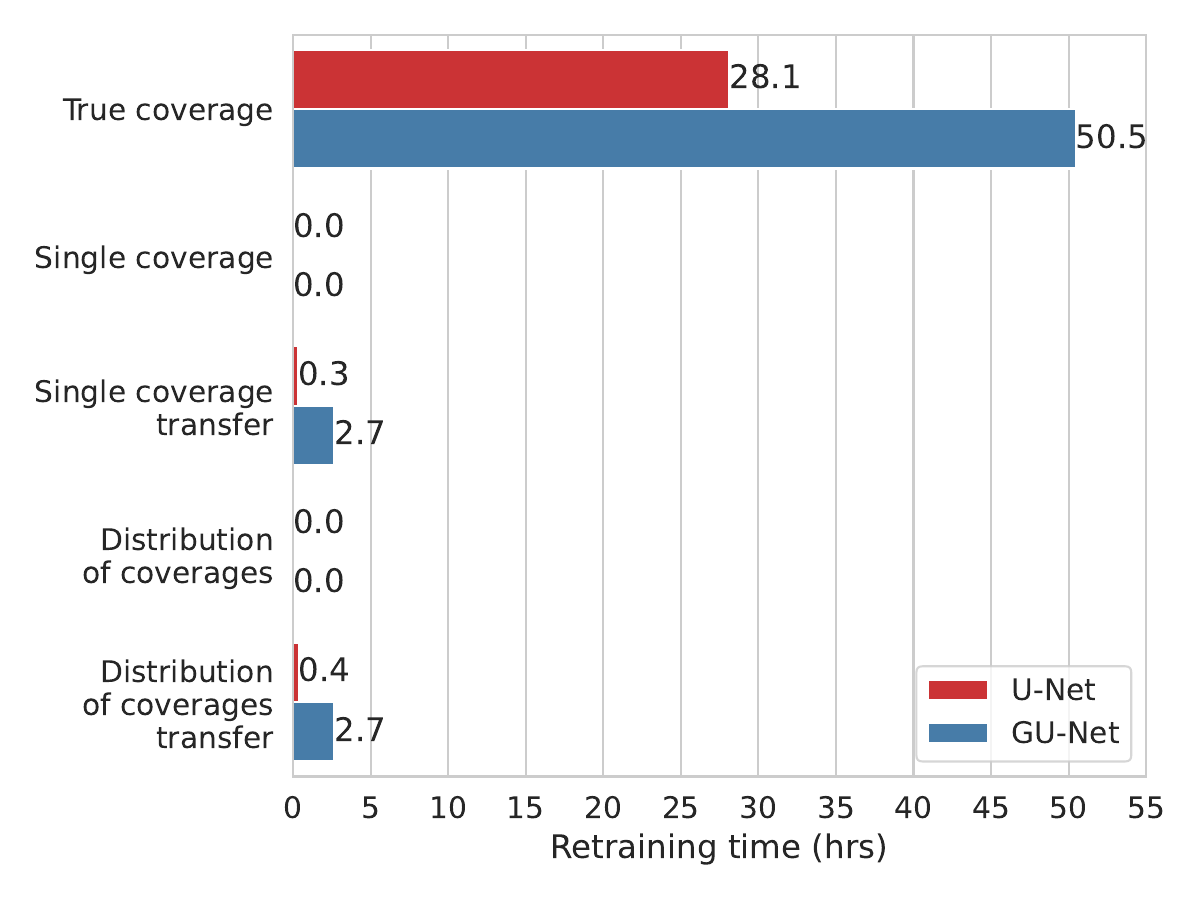}
  \caption{The time necessary for training both the U-Net and GU-Net models (on one A100 GPU with 40Gb VRAM) once visibility coverage is available. The proposed approaches reduce post-observation retraining time from 28 and 51 hours to zero hours if not fine-tuned, for the U-Net and GU-Net respectively. If transfer learning is applied, fine-tuning the models, to improve reconstruction quality retraining time is reduced to 0.3 and 2.7 hours respectively. Note that the time for training the full model (the \emph{true coverage} strategy) also contains the creation and pre-augmentation of the training data which took $\sim 24$hrs. }
  \label{fig:training_times}
\end{figure}

The main aim of the different training strategies is to reduce the time needed to (re)train a model for each observation, while still maintaining high reconstruction quality.
Figure~\ref{fig:training_times} shows the time needed to get a reconstruction network ready for image reconstruction.
The training of the full models takes more than 1 day for the U-Net (on one A100 GPU with 40Gb VRAM) and more than 2 days for the GU-Net models and are therefore not feasible to be retrained for every observation.
A large portion of this time ($\sim 24$hrs) comes from the creation and pre-augmentation of the data set using the specific \emph{uv}-coverage(s) for training.
Instead we have two strategies that require no retraining: the single coverage and distribution of coverages.
Furthermore, we have two approaches that only require a small amount of retraining in the form of transfer learning.
These transfer learned models do not require a large and augmented data set to be created and can be trained in a matter of minutes or hours.

Table~\ref{tab:reconstruction-time} shows the reconstruction time necessary for the U-Net and GU-net models once trained.
The reconstruction is dominated by the evaluations of the measurement operator to create the initial reconstruction for both models and for the calculation of the gradients and dirty images in the GU-Net.
Reconstruction time scales linearly with the number of full-scale evaluations of the operator.
The computational cost of state-of-the-art iterative reconstruction methods also scales linearly with the number of evaluations of the measurement operator as these are evaluated at each iteration.
However, these techniques typically require an order of magnitude more iterations to converge and are therefore much more computationally expensive than the proposed methods, which require only a few evaluations.

\begin{table}
  \centering
  \caption{The average reconstruction time by creating the dirty image using the pseudo-inverse, and the reconstructions using the U-Net and the GU-Net on an A100 GPU. We also detail the number of full-scale evaluations of the measurement operator and its adjoint.
  }
  \label{tab:reconstruction-time}
  \begin{tabular}{lcc} 
    \hline
    Name           & Operator evaluations & Reconstruction time (ms)      \\
    \hline
    Pseudo-inverse & $1$                  & $13.0\pm 0.4$                 \\
    U-Net          & $1$                  & $54.9 \pm 1.6$                \\
    GU-Net         & $7^*$                & $85.7 \pm 1.5$                \\
    \hline
    \multicolumn{3}{p{.9\columnwidth}}{${}^*$Refers to operator evaluations at the finest scale, which dominates the computational time of the GU-Net.} \\
  \end{tabular}
\end{table}

\subsection{Reconstruction quality}
\begin{figure*}
  \centering
  \includegraphics[height=.34\textwidth]{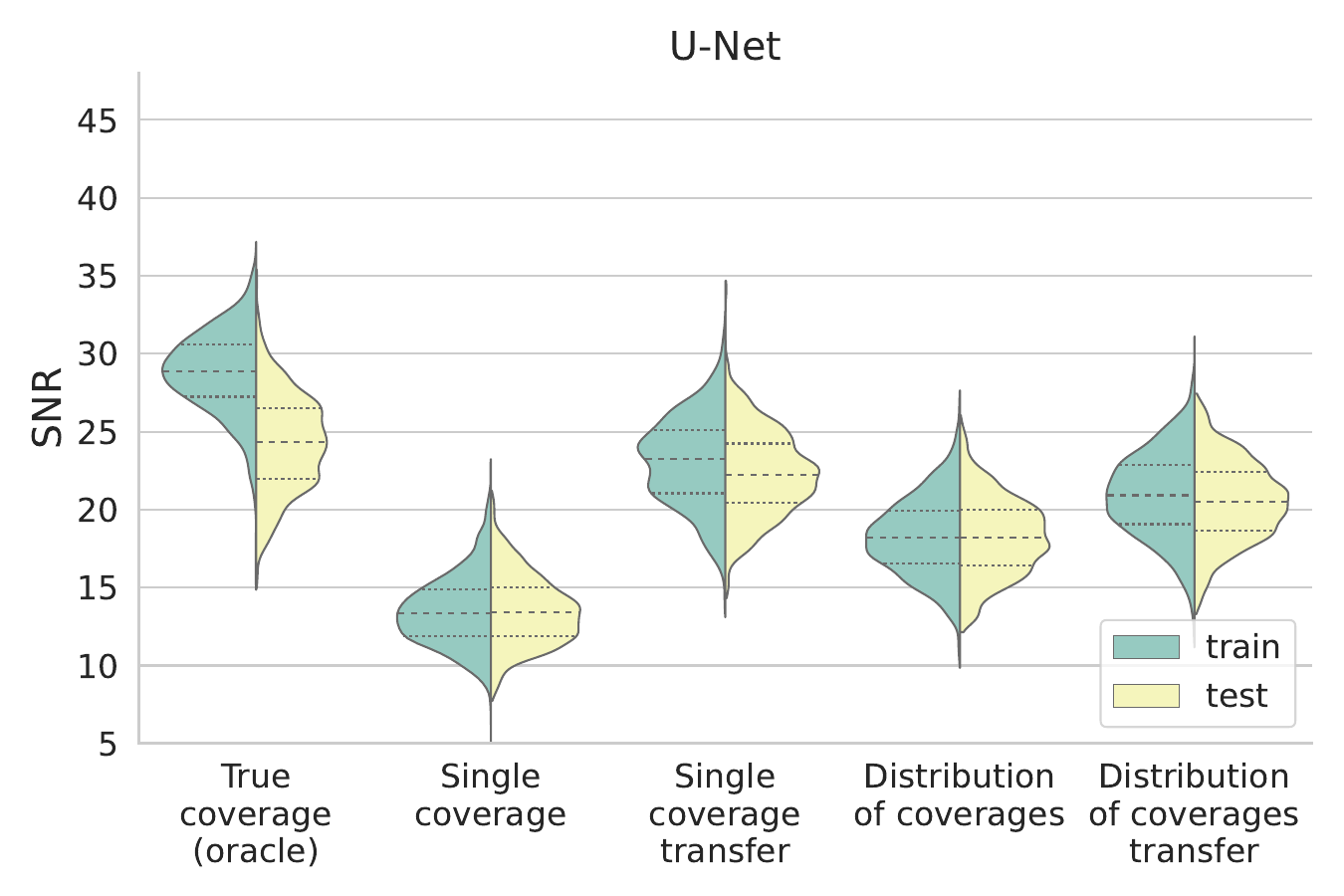}
  \includegraphics[height=.34\textwidth, trim= 2.2cm 0 0 0, clip]{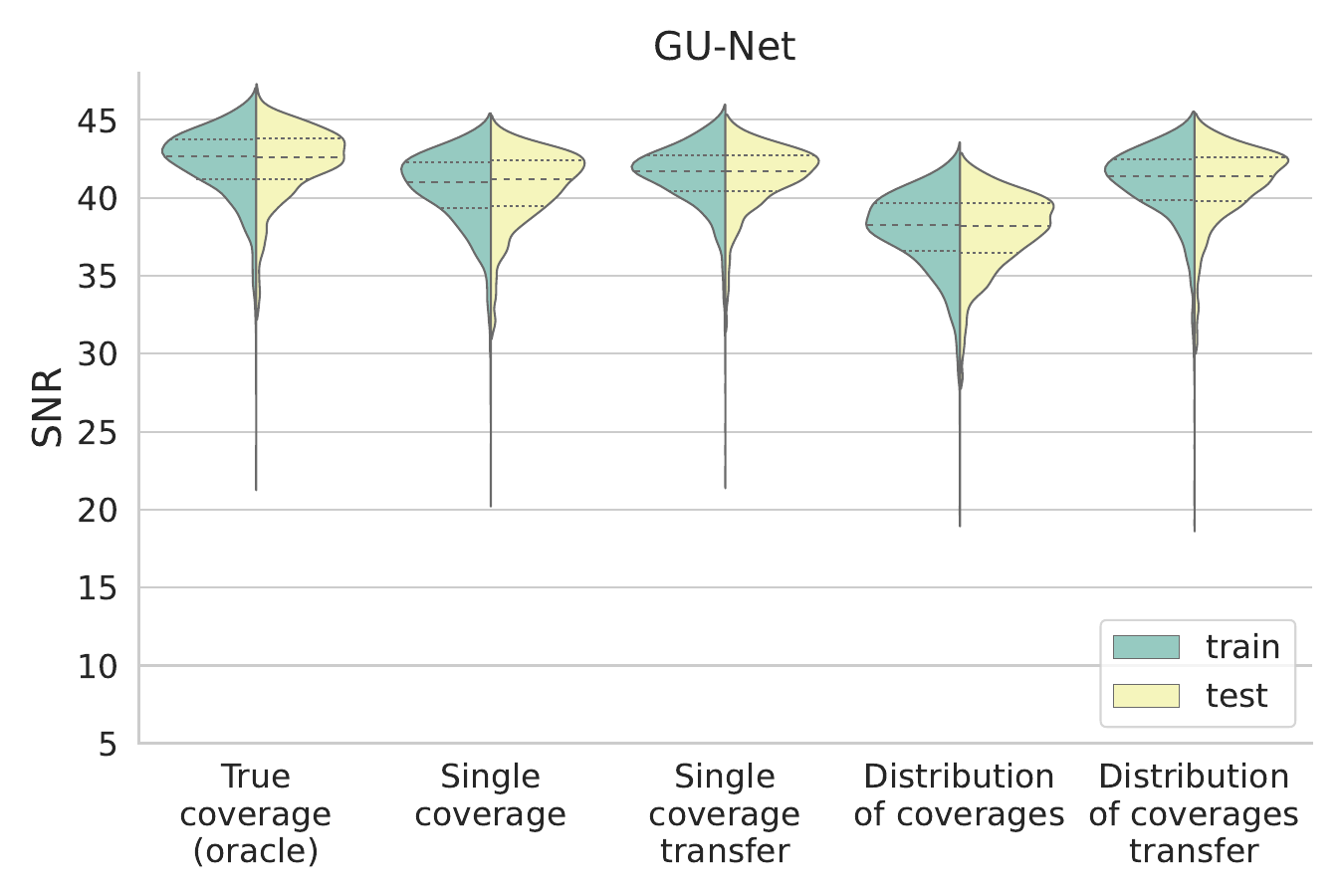}
  \caption{Reconstruction quality measured in SNR for the different training strategies for both the U-Net and the GU-Net architectures. The GU-Net models provide significantly higher reconstruction quality than the U-Net models and perform similarly across the different strategies. Without any retraining (the single and distribution of coverages strategies) these models lose only a small amount of reconstructive performance compared to fully retraining on the true coverage. Fine-tuning using transfer learning for a short time restores performance close to that of the true coverage. A similar trend across training strategies can be seen in the U-Net though the difference in performances is much larger and the transfer learning is not able to fully restore the reconstructive performance.}
  \label{fig:violin}
\end{figure*}

We assess the reconstruction quality for the five training strategies by evaluating the SNR on reconstructions of images from the train and test set.
Figure~\ref{fig:violin} shows the distribution of the reconstruction SNR for both the U-Net and the GU-Net on these simulated measurements.

The performance of the U-Net architecture is significantly worse than the GU-Net architecture, with the GU-Net architecture achieving a higher reconstruction quality for all the different training strategies, indicating that adding in measurement information in the reconstruction process significantly improves the reconstruction quality.

For the U-Net, we also notice that training the model using data from the true coverage results in a model that overfits to the training data set, as the reconstruction quality on the test set is significantly worse than the training set.
While this could be improved by using a larger training set or a training set with more variance (or by including additional methods to combat overfitting), this tendency to overfit shows that this reconstruction approach itself relies heavily on the prior information in the training set and will not generalise well to images outside of the training distribution.
Furthermore, we can see this in the performance of the U-Net trained on a single coverage, which generalises very poorly when used to reconstruct images from measurements using the true coverage.
This is not unexpected since the U-Net reconstruction only includes the measurement operator explicitly in the initial reconstruction passed to the network.
Therefore this method is biased towards the artefact distribution of the \emph{uv}-coverage used in training.
Training the U-Net using data of a distribution of coverages alleviates some of these issues as the network is able to learn the prior information for an arbitrary set of coverages with some performance loss compared to training on the actual coverage.
A minimal amount of transfer learning increases the performance for both strategies, but the performance is still worse than fully retraining a model using data simulated using the true coverage.
Longer retraining would likely further increase the performance of these models, but at some point we risk overfitting to the train set as we observed in the model trained on the true coverage.

The performance of the GU-Net is more uniform across the different strategies.
With none to little retraining we achieve significantly closer results to training using the true coverage.
The integration of the measurement operator inside the reconstruction network allows the network to learn the prior information irrespective of the coverage used.
This is reflected in the performance of the single coverage model which generalises well when used to reconstruct images from measurements using the true coverage, without any retraining.
The distribution of coverages strategy seems to generalise slightly worse to the true measurement set, which is likely due to the fact that it is trying to learn a more general prior that works for any coverage.
The discrepancy between these two strategies is directly linked to the distribution of \emph{uv}-coverages used for training.
In this case the errors induced by the coverage used for the single coverage network are similar enough to that of those induced by the true coverage resulting in good generalisation from one to the other.
However, if the distribution of telescope \emph{uv}-coverages has a significantly larger variance it would be expected that the single coverage model would not generalise as well to the true coverage and the distribution of coverages model might be preferred.
Despite this, the performance of both models is improved using a small amount of transfer learning, bringing the reconstruction quality close to that of the oracle model trained using the true coverage.

\begin{figure*}
  \centering
  \includegraphics[width=\textwidth, trim= 12.5cm 3.75cm 8.5cm 1.3cm, clip]{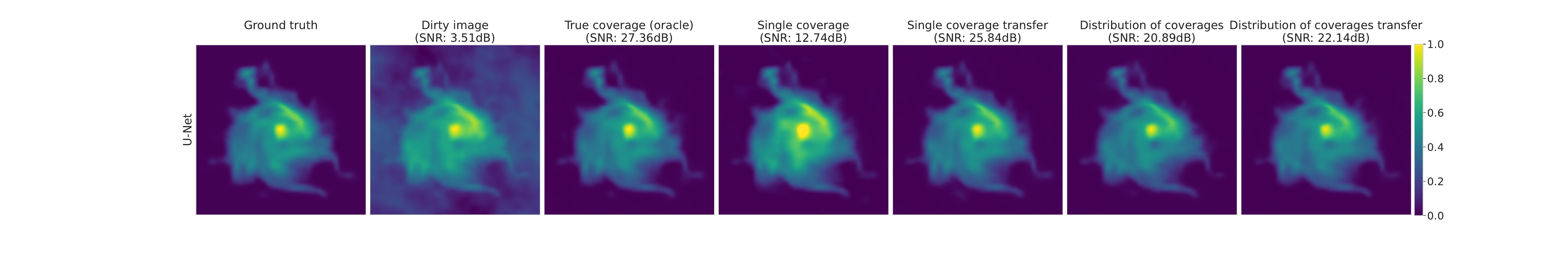}
  \includegraphics[width=\textwidth, trim= 12.5cm 3.75cm 8.5cm 2.0cm, clip]{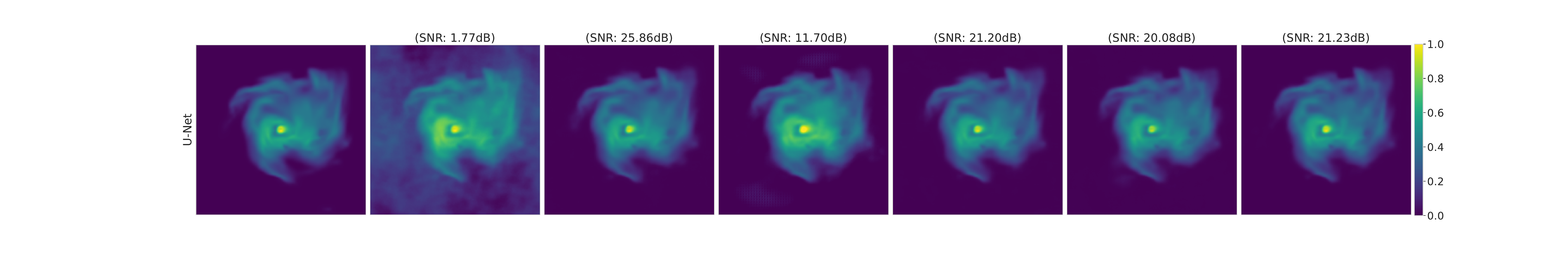}
  \includegraphics[width=\textwidth, trim= 12.5cm 3.75cm 8.5cm 2.0cm, clip]{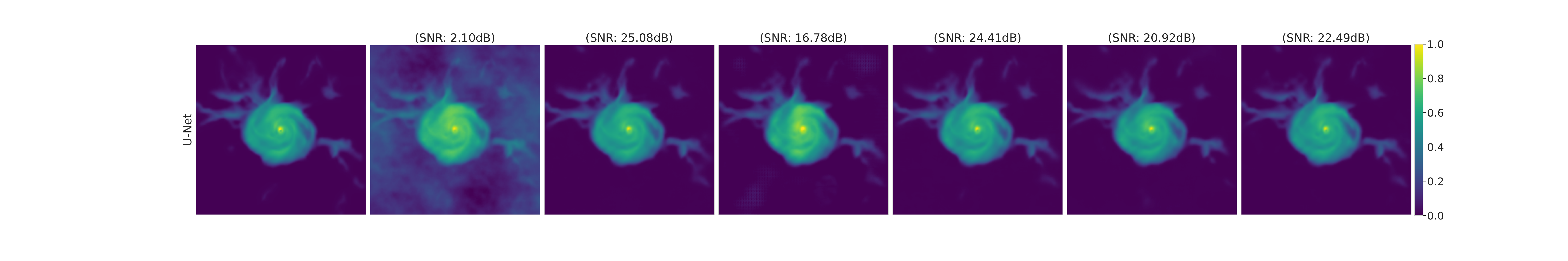}
  \includegraphics[width=\textwidth, trim= 12.5cm 3.75cm 8.5cm 2.0cm, clip]{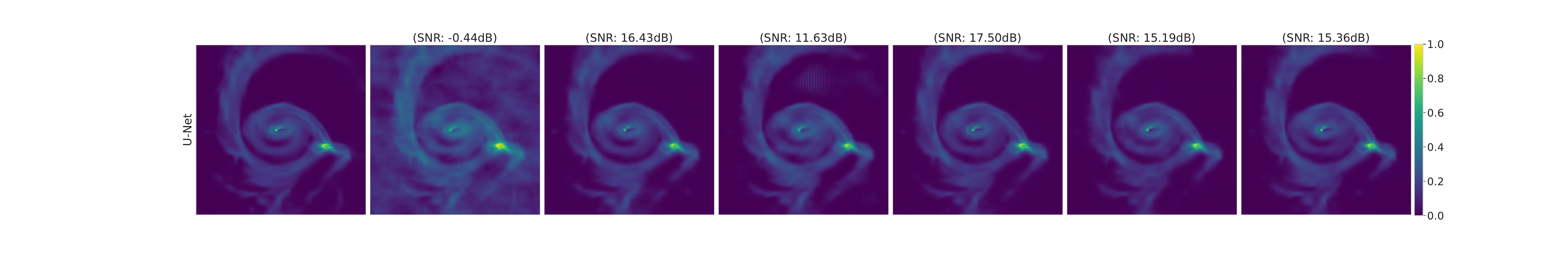}
  \includegraphics[width=\textwidth, trim= 12.5cm 3.75cm 8.5cm 2.0cm, clip]{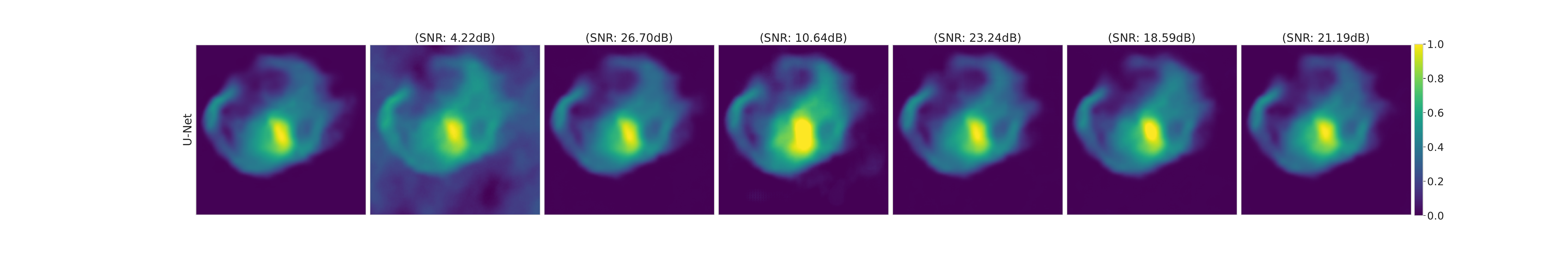}

  \caption{Reconstructions using the trained variants of the U-Net from simulated measurements of galaxy images using the true \emph{uv}-coverage drawn from a Guassian random distribution. Without any retraining, the single coverage and distribution of coverages strategies, struggle to reproduce the reconstruction quality achieved by fully retraining the models (the true coverage strategy), as indicated by the reconstruction SNR values. In particular the single coverage strategy performs poorly, sometimes even showing hallucinations of faint structure (e.g. the fourth example top centre). Using transfer learning improves the SNR of the reconstructions for both strategies, though performance is still lacking compared to the true coverage, oracle strategy. }
  \label{fig:examples_TNG_UNet}
\end{figure*}

Figure~\ref{fig:examples_TNG_UNet} shows example reconstructions using the U-Net for the different training strategies.
While the reconstructions are visually similar the reconstruction quality as indicated by the SNR is significantly different and reflects the results in Figure~\ref{fig:violin}.
The difference in the reconstruction strategies can be seen in their accuracy of predicting the diffuse structures of the images as well as correctly reconstructing the intensity of the brighter features.
Besides that, we see that the model using the single coverage strategy produces reconstructions that contain hallucinations of a diffuse structure that are not found in the ground truth (most notably in the fourth and fifth example reconstructions).

\begin{figure*}
  \centering
  \includegraphics[width=\textwidth, trim= 12.5cm 3.75cm 8.5cm 1.3cm,  clip]{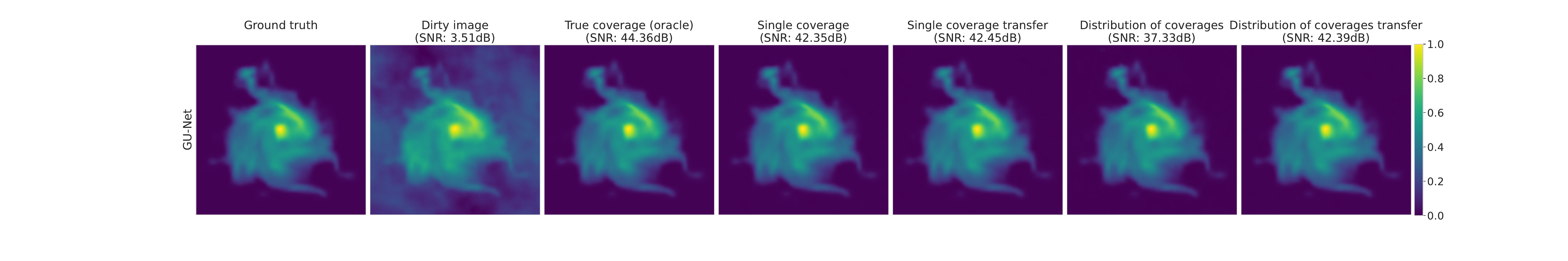}
  \includegraphics[width=\textwidth, trim= 12.5cm 3.75cm 8.5cm 2.0cm, clip]{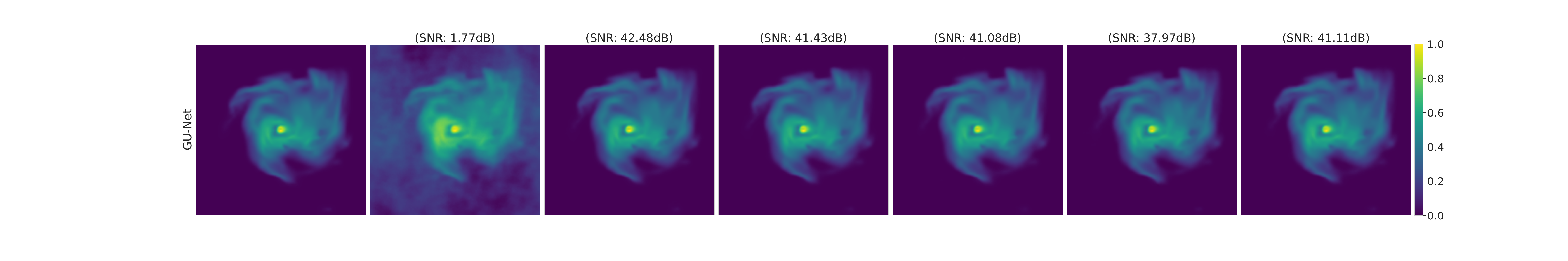}
  \includegraphics[width=\textwidth, trim= 12.5cm 3.75cm 8.5cm 2.0cm, clip]{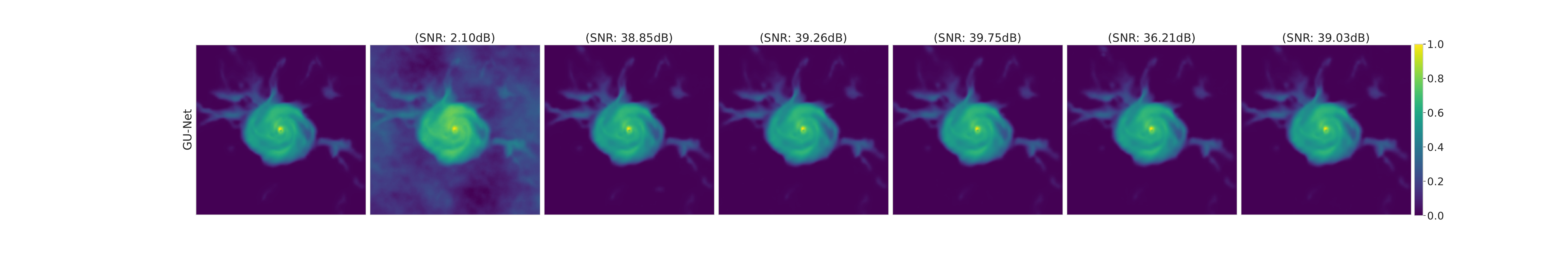}
  \includegraphics[width=\textwidth, trim= 12.5cm 3.75cm 8.5cm 2.0cm, clip]{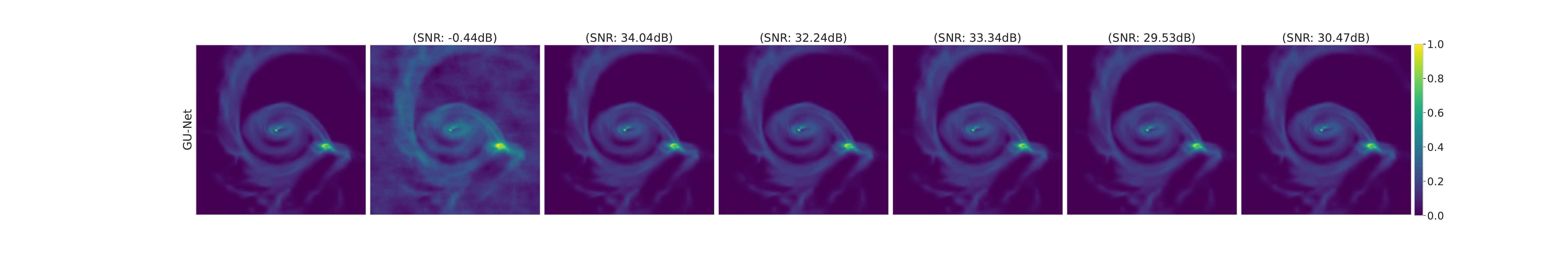}
  \includegraphics[width=\textwidth, trim= 12.5cm 3.75cm 8.5cm 2.0cm, clip]{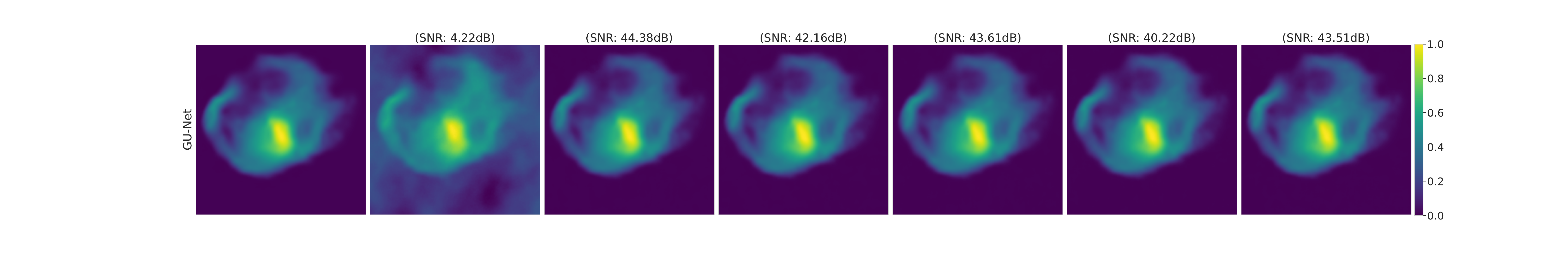}
  \caption{Reconstructions using the trained variants of the GU-Net from simulated measurements of galaxy images using the true \emph{uv}-coverage  drawn from a Guassian random distribution. The strategies without retraining, the single coverage and distribution of coverages, generalise well to the unseen true coverage at only a minor loss in SNR compared to the oracle model trained on the true coverage. A small amount of fine-tuning using transfer learning brings the performance of these strategies closer to the true coverage, oracle model. }
  \label{fig:examples_TNG_GUNet}
\end{figure*}

In Figure~\ref{fig:examples_TNG_GUNet} we see the reconstructions using the GU-Net for the different training strategies.
The different strategies produce visually similar looking reconstructions with high reconstruction quality.
While the network using a distribution of coverages for training performs slightly worse than the other strategies, using a small amount of transfer learning improves the performance significantly to match that of the other strategies.

\subsection{Reconstruction quality for increased dynamic range}\label{sec:high-dynamic-range}
While the proposed methods are trained on images of simulated galaxies, the resulting networks should also be tested on more realistic radio observations.
Therefore, to test the generalisation power of the different training strategies, we evaluate the performance of the models on an out-of-sample image of 30 Doradus.
The dynamic range of this image is larger than the simulated images ($\sim 600$) and we therefore also evaluate the reconstructions on a logarithmic scale.
While this image has a higher dynamic range than the simulated images, it is not as high as the dynamic range of real radio observations (which can be in the order of $10^6$).

\begin{figure*}
  \centering
  \includegraphics[width=\textwidth, trim= 12.5cm 3cm 8.2cm 1.3cm,  clip]{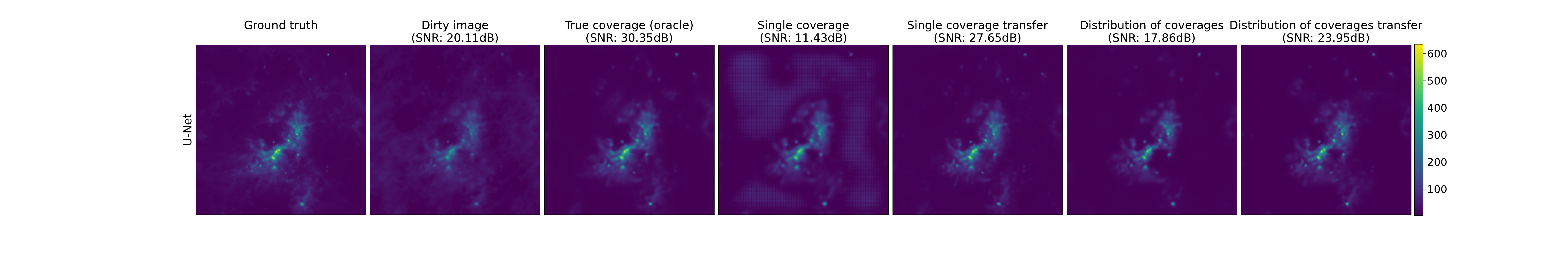}
  \includegraphics[width=\textwidth, trim= 12.5cm 3cm 8.2cm 2.2cm, clip]{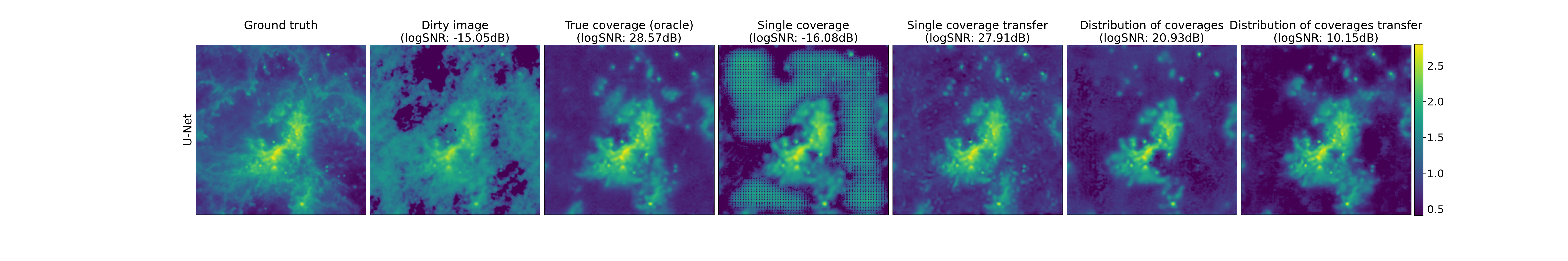}
  \caption{Reconstructions using the trained variants of the U-Net on simulated measurements, using the true \emph{uv}-coverage drawn from a Gaussian random distribution, of an out-of-sample image of 30 Doradus, displayed on a linear scale (top) and a logarithmic scale (bottom). The strategies that are not fine-tuned on data from the true coverage perform significantly worse in reconstructing this out-of-sample image. Using transfer learning from the single coverage strategy brings the reconstructive SNR closer to fully retraining the model on the true coverage. All U-Net reconstructions struggle to accurately reconstruct the larger dynamic range of this image as indicated by the logarithmically scaled images and the logSNR, with the single coverage strategy producing large artefacts. }
  \label{fig:examples_radio_UNet_30Dor}
\end{figure*}

Figure~\ref{fig:examples_radio_UNet_30Dor} shows the reconstructions using the U-Net for the different training strategies.
The models that are not trained or fine-tuned using data from the true coverage perform significantly worse for this post-processing approach, whereas we see that fine-tuning the models produces a higher quality of reconstructions.
Nonetheless, the models are not able to reconstruct the dynamic range of the image correctly, with the reconstructions showing large amounts of noise and artefacts.
This lack of performance traces back to the fact that the models rely heavily on the prior information in the training set and are not able to generalise well to new data with a higher dynamic range.

\begin{figure*}
  \centering
  \includegraphics[width=\textwidth, trim= 12.5cm 3cm 8.2cm 1.3cm,  clip]{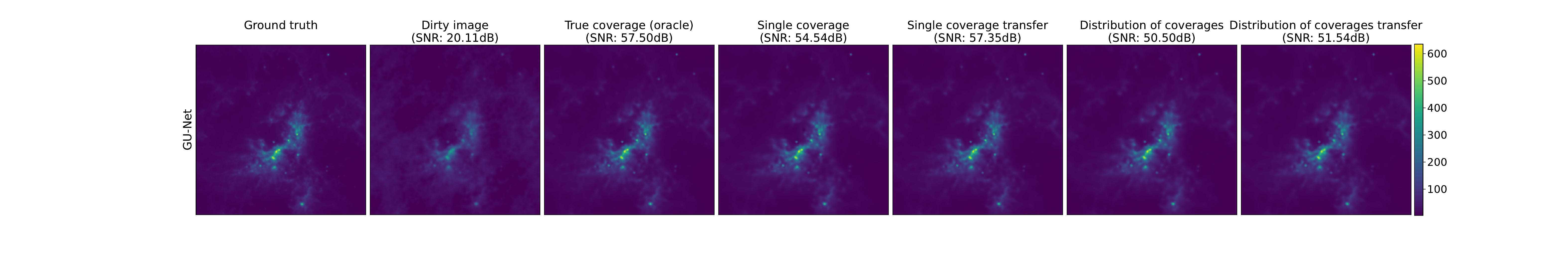}
  \includegraphics[width=\textwidth, trim= 12.5cm 3cm 8.2cm 2.2cm, clip]{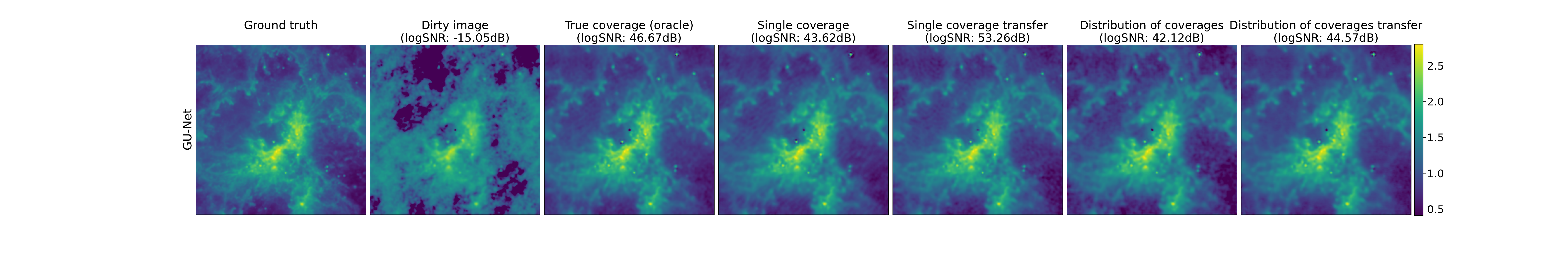}
  \caption{Reconstructions using the trained variants of the GU-Net on simulated measurements, using the true \emph{uv}-coverage drawn from a Gaussian random distribution, of an out-of-sample image of 30 Doradus, displayed on a linear scale (top) and a logarithmic scale (bottom). The GU-Net is able to reconstruct the larger dynamic range of the image across the different strategies as indicated by both the high SNR and logSNR of the reconstructions. Training the model on a single coverage and then fine-tuning it using transfer learning provides reconstruction quality close to that of the model that is fully trained on simulated data from the true coverage. }
  \label{fig:examples_radio_GUNet_30Dor}
\end{figure*}

When looking at the reconstructions from the same measurements using the GU-Net in Figure~\ref{fig:examples_radio_GUNet_30Dor}, we see that the models are able to produce high quality reconstructions with a larger dynamic range.
Similar to on the simulated images, the methods not fine-tuned on data from the true coverage performs slightly worse but this difference is easily mitigated by fine-tuning the models using transfer learning, yielding reconstructions that are of the same reconstruction quality as fully retraining the model.
The explicit inclusion of the measurement operator in the reconstruction process allows the network to learn the prior information irrespective of the coverage used, which is why the GU-Net is able to generalise much better to out-of-distribution data as well as generalising to different \emph{uv}-coverages.
This generalisablity makes it more suitable for application to real radio observations, since it is able to correct for differences between the training data set and the actual observations.

\subsection{Transfer learning to MeerKAT \emph{uv}-coverage}\label{sec:transfer-meerkat}
\begin{figure}
  \centering
  \includegraphics[width=0.8\columnwidth, trim = 0.3cm 0.3cm 0.5cm 1.5cm, clip]{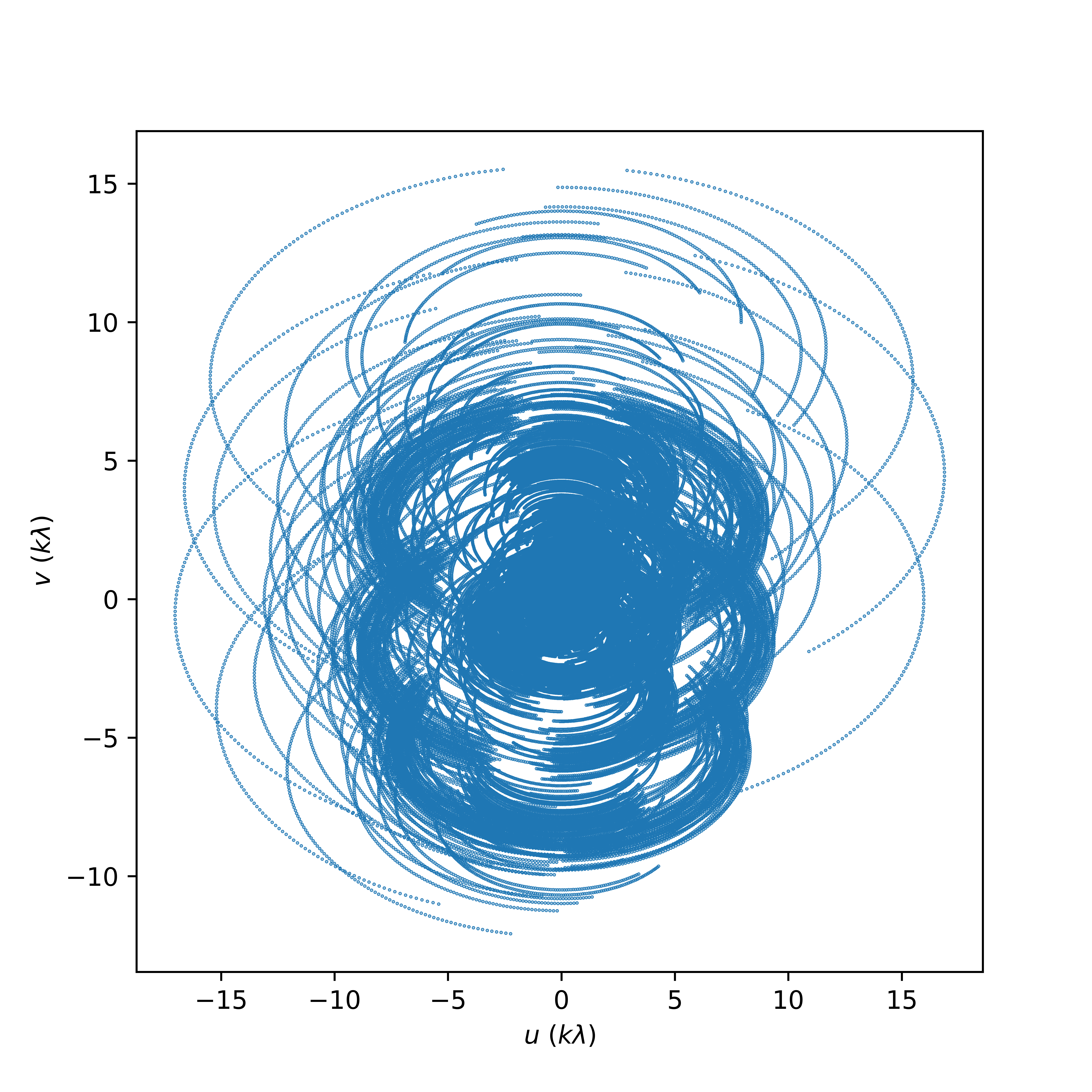}
  \caption{The \emph{uv}-coverage of an observation with the MeerKAT telescope.}
  \label{fig:meerkat-uv}
\end{figure}

\begin{figure*}
  \centering
  \includegraphics[width=\textwidth, trim= 5cm 2.7cm 2.2cm 1.5cm,  clip]{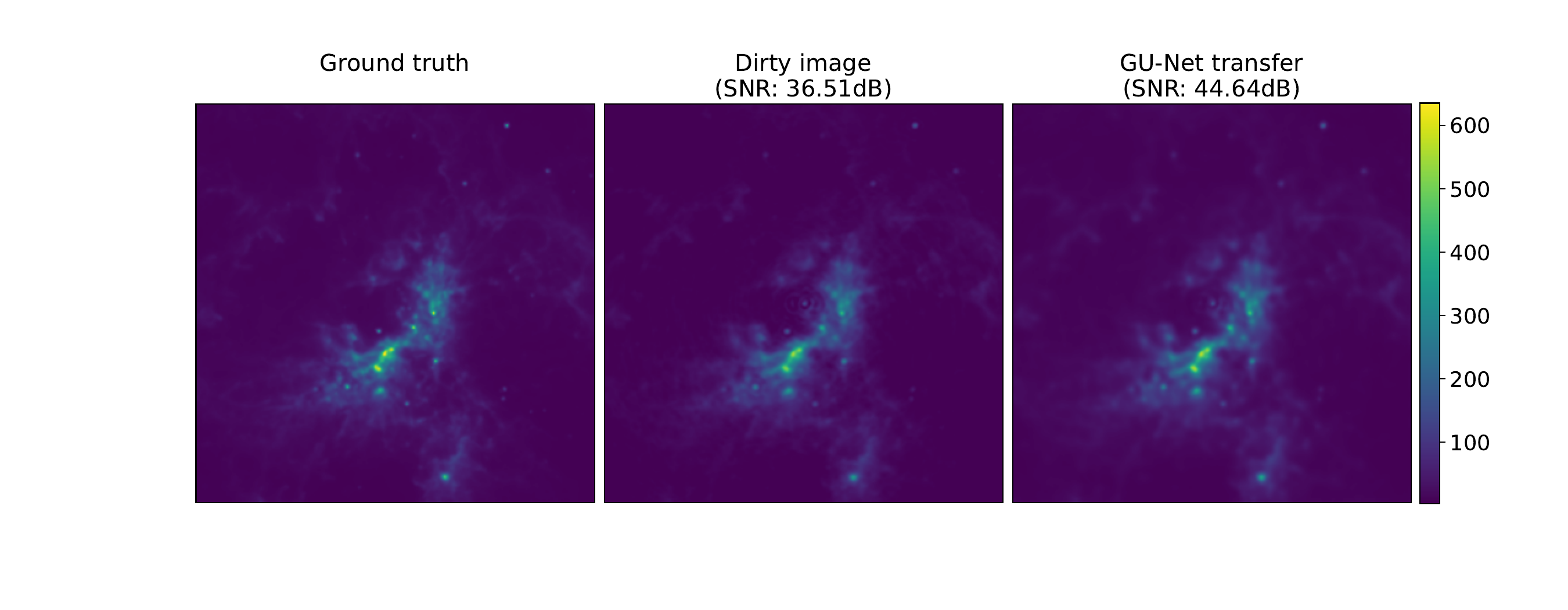}
  \includegraphics[width=\textwidth, trim= 5cm 2.7cm 2.2cm 2.3cm, clip]{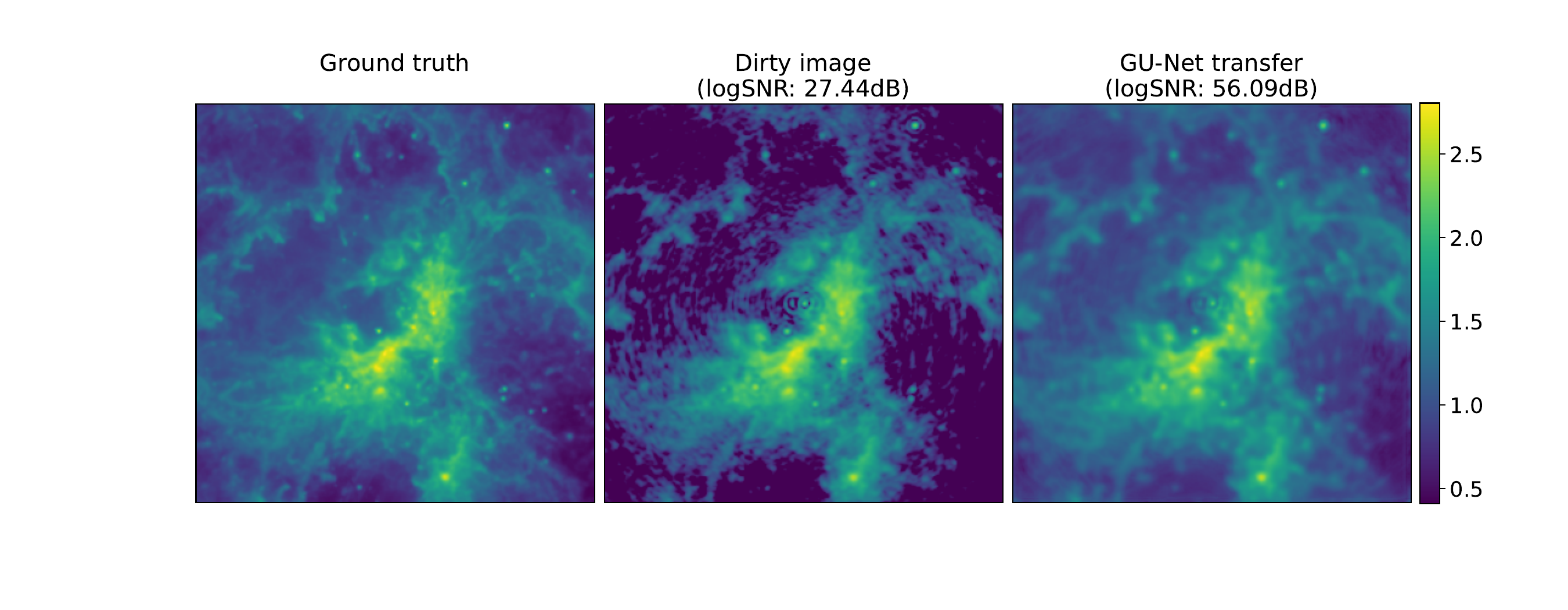}
  \caption{Reconstructions using a fine-tuned variant of the GU-Net on simulated measurements on an out-of-sample image of 30 Doradus using the MeerKAT \emph{uv}-coverage as displayed in Figure~\ref{fig:meerkat-uv}. The ground truth, dirty image and GU-Net reconstruction are displayed on a linear scale (top) and a logarithmic scale (bottom). Using this fine-tuned model we achieve a high quality reconstruction that also reproduces the larger dynamic range of the radio observation, using only minimal amount of re-training.  }
  \label{fig:examples_meerkat_GUNet_30Dor}
\end{figure*}

Finally, to investigate whether these reconstruction networks generalise from the Gaussian visibility coverages to realistic \emph{uv}-coverages, we use the same models to reconstruct from simulated measurements using a \emph{uv}-coverage from the MeerKAT telescope with $241920$ visibilities as seen in Figure~\ref{fig:meerkat-uv}.
In order to do this we fine-tune the GU-Net model trained on the single coverage by training it on a small data set created using the MeerKAT coverage.
For the creation of the dirty image and the gradient operations inside the GU-Net model, uniform measurement weighting is used. 
The model is trained for 50 epochs and the creation of the data took $\sim 2$ hours and the fine-tuning of the model took $\sim 5$ hours on an A100 GPU with 40GB VRAM.
For this task we only consider the GU-Net architecture as the U-Net models do not generalise well to larger dynamic range images as demonstrated in Section~\ref{sec:high-dynamic-range}.
The fine-tuned GU-Net is then used to reconstruct an out-of-sample image of 30 Doradus using the MeerKAT coverage.
The results are shown in Figure~\ref{fig:examples_meerkat_GUNet_30Dor}.

We see that the transfer learned model is capable of reconstructing images with a larger dynamic range, despite not being trained on any data with a similar dynamic range.
Since the number of measurements in the MeerKAT coverage is larger than that of the Gaussian coverage ($\sim 7.4\times$ more measurements), the reconstruction time is longer ($194.6 \pm 3.1$ms; $\sim 2.3\times$  longer) than for the simulated measurements and the training takes longer.
\section{Conclusions}\label{sec:conclusion}
Since the \emph{uv}-coverage of every radio interferometric observation depends on factors such as the pointing direction of the telescope and the rotation of the Earth, it is essential to provide reconstruction techniques that generalise across varying visibility coverages. 
To this purpose we have proposed several training strategies for learned reconstruction methods that provide such robustness with minimal to no retraining of the reconstruction networks.

First of all, we conclude that methods that do not include the measurement operator in the reconstruction network, such as learned post-processing methods based on a U-Net model, are not robust to changes in the \emph{uv}-coverage and are highly reliant on learning features that are part of the training data. 
While these methods are the most computationally efficient reconstruction approaches, their reconstructive performance is significantly worse than methods that do include the measurement operator, even when fully (re)trained or fine-tuned to the true \emph{uv}-coverage, due to overfitting.
Furthermore, by looking at the performance of this post-processing approach on more realistic radio interferometric observations, we have shown that these methods struggle to capture the larger dynamic range of these images across the different training strategies. 
This indicates that these methods are limited by the prior information in their training data, which is concerning in the context of real data where the true image is unknown.

On the other hand, we have shown that methods that include the measurement operator in the reconstruction network, such as the GU-Net model \citep{marsLearnedInterferometricImaging2023b}, maintain a high reconstruction quality across different \emph{uv}-coverages with minimal to no need for retraining. 
In case of a Gaussian distribution of \emph{uv}-coverages, training on a single different \emph{uv}-coverages provides us with a network which suffers only a minor decrease in reconstruction quality on other \emph{uv}-coverages of the same distribution with no retraining. 
With only a small amount of fine-tuning on the true \emph{uv}-coverage, we come close to the reconstruction quality of a fully retrained network. 
This indicates that these methods are able to learn a more general representation of the prior information in the training data, irrespective of the \emph{uv}-coverage.

Furthermore, we have shown that when applying the GU-Net reconstruction method to measurements outside of the training domain, such as a more realistic radio observation with a larger dynamic range ($\sim600$), we retain this reconstruction quality, indicative of the approach being significantly less reliant on the prior information in the training data and able to generalise across different types of data. 

Finally, we have demonstrated that this unrolled iterative reconstruction method based on the GU-Net can also be used to reconstruct images from more realistic \emph{uv}-coverage distributions, such as the MeerKAT \emph{uv}-coverage, with high quality. 
While we show reconstruction models fine-tuned from Gaussian \emph{uv}-coverage to the realistic MeerKAT \emph{uv}-coverage, we recommend that future work investigate networks trained on a distribution of \emph{uv}-coverages that is more representative of the true \emph{uv}-coverage distribution, as this might provide us with a better starting point for fine-tuning the networks to the true \emph{uv}-coverage.

To extend these method to and test them on real observations, several challenges need to be addressed.
For accurate modelling of the measurements, there is need for a version of the radio interferometric measurement operator in a fully automatically differentiable framework that includes wide-field effects, such as w-stacking and w-projection methods.  
However, using the full measurement operator during training for large \emph{uv}-coverages would be highly computationally demanding and require a huge amount of memory on the GPU(s) to train for an SKA-sized \emph{uv}-coverage. 
In order to address this, we propose to investigate the use of approximate measurement operators in the reconstruction network by using a convolution with the PSF to approximate the measurement operator as a way to reduce the computational overhead of changing the coverage during training, as has been investigated for proximal optimisation methods \citep{besterPracticalPreconditionerWidefield2021}. 
This approximation approach has also been used in \citet{aghabiglouR2D2DeepNeural2024} to train a series of residual networks to reconstruct the image from the measurements.
While this approximation does not model wide-field effects when the baselines are non co-planar, this would allow us to train on a distribution of realistic \emph{uv}-coverages without the need to change the coverage during training, which would be essential in order to train these networks. 
However, the difference in performance between using the full measurement operator and the approximate operator in learned methods should be investigated further to assess how much information is lost.

While we show that the GU-Net reconstruction models are able to generalise to a larger dynamic range ($\sim600$) than the training data, we note that this dynamic range is still orders of magnitude lower than the dynamic range of real observations (which can be $\sim10^6$).
In order to extend these methods to real observations, more extensive investigation should be performed with regards to creating a representative training data set, and thus prior information, that includes the high dynamic range and bright point sources that are present in real observations, as well as varying noise levels, beam errors, calibration errors and other errors that might be present in the actual data.

An interesting avenue for future work would be to investigate not only the fine tuning on the true \emph{uv}-coverage, but also the performance of transfer learning on real data. 
In \citet{marsLearnedInterferometricImaging2023b} we have shown that transfer learning can be used to improve the performance of our reconstruction methods trained on a general image data set, by using a small amount of domain specific data. 
This approach could bridge the gap between a larger data set of simulated data and a small data set of real data.

In conclusion, we have presented a comprehensive study of the robustness of learned reconstruction methods to changes in the \emph{uv}-coverage, and have shown that methods that include the measurement operator in the reconstruction network are able to generalise across different \emph{uv}-coverages with minimal to no need for retraining, while maintaining a high reconstruction quality. 
We have shown that these methods are able to learn a generalised representation of the prior information in the training data, irrespective of the \emph{uv}-coverage, and are able to generalise across different types of data, which is essential when looking to apply these methods to real data.

\section*{Acknowledgements}
We thank Tariq Blecher for interesting discussions.
Matthijs Mars is supported by the UCL Centre for Doctoral Training in Data Intensive Science (STFC Training grant ST/P006736/1).
This work is also partially supported by EPSRC (grant number EP/W007673/1).
The authors acknowledge the use of the UCL Myriad High Performance Computing Facility (Myriad@UCL), and associated support services, in the completion of this work.
This work also used computing equipment funded by the Research Capital Investment Fund (RCIF) provided by UKRI, and partially funded by the UCL Cosmoparticle Initiative.

\section*{Data Availability}
The code used to generate the results in this paper is available at \url{https://github.com/astro-informatics/LeIA}. The data used in this paper is available upon reasonable request to the corresponding author.

\newpage



\bibliographystyle{mnras}
\bibliography{Bibliography/lib.bib} 








\bsp	
\label{lastpage}
\end{document}